\documentclass[12pt]{article}
\usepackage{latexsym}
\makeatletter
\@addtoreset{equation}{section}

\makeatother

\begin{document}

\newcommand{\be}{\begin{equation}}
\newcommand{\ee}{\end{equation}}
\newcommand{\<}{\langle}
\renewcommand{\>}{\rangle}
\newcommand{\reff}[1]{(\ref{#1})}

\title{A free action for pions as quark composites} 
\author{
  { Sergio Caracciolo}             \\[-0.2cm]
  {\small\it Scuola Normale Superiore and INFN -- Sezione di Pisa}  \\[-0.2cm]
  {\small\it I-56100 Pisa, ITALIA}          \\[-0.2cm]
  {\small Internet: {\tt Sergio.Caracciolo@sns.it}}     \\[-0.2cm]
  \\[-0.1cm]  \and
  { Fabrizio Palumbo~\thanks{This work has been partially 
  supported by EEC under TMR contract ERB FMRX-CT96-0045}}             \\[-0.2cm]
  {\small\it INFN -- Laboratori Nazionali di Frascati}  \\[-0.2cm]
  {\small\it P.~O.~Box 13, I-00044 Frascati, ITALIA}          \\[-0.2cm]
  {\small Internet: {\tt palumbof@lnf.infn.it}}     
   }
\date{September 18, 1997}
\maketitle

\thispagestyle{empty}   


\begin{abstract}
In the framework of an approach to bosonization based on the use of 
fermionic composites as fundamental variables, a quadratic
action in even Grassmann variables with the quantum numbers of the pions 
 has been constructed. It includes the $\sigma$-field in order to be invariant
under
$[SU(2)]_L\otimes[SU(2)]_R$ transformations over the quarks. This action
 exhibits the Goldstone phenomenon reducing its symmetry to
the O(3) isospin invariance.
A squared mass for the pions is generated according to PCAC.

The model has been investigated in the Stratonovitch--Hubbard representation,
in which form it is reminiscent of the Gell-Mann-L\'{e}vy model.
By the 
 saddle point method a renormalizable expansion in inverse powers of the
 index of nilpotency of the mesonic fields (which is 24), is generated.
The way it might be used in a new perturbative approach to QCD is outlined.
 \end{abstract}

\clearpage

\section{Introduction}\label{introduction}

Recently it has been suggested that the correlation functions of fermionic
 composites
can be evaluated by a change of variables whereby the
composites themselves are assumed as integration variables.  Although the
 constituents cannot be expressed in terms of the
composites, we can talk about a generalized change of variables in
a precise sense~\cite{Palu}.

In the context of QCD the motivation is to set up
a perturbative expansion in terms of the hadrons instead of the usual one in
terms of the quarks. A prerequisite for this goal is the existence of a free action
for the composites. Surprisingly enough, {\it for trilinear composites like the barions the
free action turns out to be the Dirac one }. 
This is due to the fact that if the
trilinear composites are chosen in the proper way, their integral is identical to the
Berezin integral over the constituents~\cite{Palu}. 

The integral over even composites instead does not reduce in general 
to a Berezin or to an ordinary integral, and 
therefore there is no reason why their free action should resemble that of
free bosons or fermions.
Indeed we remind the reader that Grassmann variables $\phi$ are characterized by
the index of nilpotency, which is the smallest integer $\Omega$ such that 
\begin{equation}
   \phi^n=0, \; \hbox{for} \;n>\Omega.
\end{equation} 
The index of odd composites is always 1 (and this is related to the simplicity of
their action), while the index of even composites depends on their structure: For the
pions, for instance, it is 24. The action of even composites is therefore expected
to be, on general grounds, a polynomial of degree
$\Omega$, and this expectation is confirmed in the cases where it has been derived.
There are
some results in the strong coupling limit of lattice QCD~\cite{Kawa},
(also at finite density~\cite{Barbour}) because in that limit the
composites appear as the natural starting point of an expansion. 
In a similar spirit in numerical simulations there have been efforts to replace the 
direct evaluation  of the fermionic determinants by the so-called 
{\em monomer-dimer expansion}~\cite{Rossi}.

All these actions are defined on a lattice and have a common feature : They are
static in the formal continuum limit. Moreover, if the Grassmann field is replaced
by a true bosonic field, they are unbounded from below.

 Our first reaction
against such a result was to attribute it to the strong coupling regime, which is
generally thought to be intrinsically far from the continuum. But this explanation
has been ruled out by the investigation of
 a simple quadratic action for a Grassmann scalar of index 1
~\cite{Palu94}. In this work there was no attempt to relate such an action to that of
constituents, but only to see whether it was possible to obtain a sensible propagator in the
continuum limit. The result was, indeed, that the action should be static 
(in this  limit) and negative definite, in which case it is possible to obtain
for the propagator a random-walk representation. 

To get further insight it was necessary to obtain in an exact way a free
action for composites starting from that of the constituents. 
For this reason  a simple nonrelativistic  
model was considered, the so-called pairing model, where the program of the use of even
Grassmann variables for the description of composites has been successfully
completed~\cite{Barb}. Here the composite is a Cooper pair which has index $\Omega=
j+{1 \over 2}$, $j$ being the angular momentum of the fermions involved ( see 
Sec. 2 ). Again the action for these composites is a polynomial of degree $\Omega$
with the properties described above, which where then conjectured to characterize all
even composites.

While this result confirms the specific features  previously
found, it is rather discouraging, because a derivation of the composite action
along the lines of the pairing model appears to be exceedingly difficult in
relativistic field theories, and in any case the resulting action too difficult to deal
with. Its complexity, however, appears due  to the fact that it
must reproduce the two-point functions of all the nonvanishing powers of  the composites
\begin{equation}
  \<[\phi^*(x) \phi(y)]^n\>,\;\;\hbox{for} \; n \leq 
  \Omega.\label{sopra}
\end{equation}
 But in Particle Physics we are essentially interested
in the correlation functions of the first power of the composites 
(namely with $n=1$ in \reff{sopra}, to be referred to in the sequel 
as the $n=1$ correlation functions), and under this restriction the
effective action might be much simpler. In Sec.~\ref{sec2}, in fact, we
investigate whether a quadratic action can yield these two-point functions for the
pairing model, and we find an affirmative answer. 

We can then explore a short
cut in the determination of an effective action for even composites,  based on
the fact that such an action, due to its high dimension, can be added to
that of the constituents by only altering by a finite amount the renormalization  group
parameters. One should proceed in the following way:  first determine, if
it exists, a quadratic free action for the mesons, and then  check in a perturbative
calculation if such an action can absorb, by appropriate renormalizations, the
contributions from the free action of the constituents. The present paper is devoted to the
first part of this program, but our results can also be regarded, in a more modest
perspective, just as the construction of a model.

In   Sec.~\ref{sec3}
we consider on a lattice a quadratic action for the  pions with the general features discussed above. We
must mention, however, that we have also found a quite different action of composites
which yields a Klein-Gordon propagator, but it gives rise to a nonrenormalizable expansion.

The propagator with an action quadratic in even Grassmann variables
{\it is not in general the inverse of the wave operator}, and it is not easy to work out.
Therefore by using the Stratonovitch--Hubbard representation~\cite{Hub} we
replace the Grassmann variables by ordinary bosonic ones, at the price of dealing
with an action which is no longer polynomial. We investigate its properties by means
of the saddle point method, which generates a $ 1 / {\Omega}$ expansion which,
in view of the rather high value of $\Omega$ for the pion,
appears quite reasonable. 

Furthermore, it is possible to choose the parameters in such a way that a free
action, with Klein-Gordon propagators, arises in the continuum limit.

The action we have studied contains the pions as well as the $\sigma$ 
fields, and it is $[SU(2)]_V \otimes [SU(2)]_A $ invariant.
This symmetry is broken
spontaneously to the $O(3)$ isospin symmetry, with massless pions and 
the expectation value of the $\sigma$ field as order parameter. 
In the presence of an explicit breaking linear in $\sigma$, 
the pions acquire a squared mass 
proportional to the breaking parameter. Results of this kind have already
been derived in many ways, but what is new here is that
they are produced by a simple quadratic action of composite fields (which can 
hopefully be used as the
unperturbed term in a new perturbative approach to QCD).

Before ending this Introduction we notice that the formalism we are going
to develop finds its easiest application in the cases where the action of the composites need
not be derived from that of the constituents, like in models where composite fields are
introduced since the beginning.
The present investigation was indeed originally motivated by such a model of
composite gauge fields~\cite{Palu}. Another application has been done in a recent work where the
possibility has been considered that the quarks might have lower scaling 
dimensions~\cite{Palumbo}. This possibility  is excluded for the leptons by the request of
 unitarity, but
such a request is much less stringent for the quarks because of confinement. 
If we assume for the quarks scaling dimension $1/2$, for instance, the hadrons
acquire  naturally the
appropriate dimensions, so that their free action has dimension four 
and can be introduced since the very
beginning in the fundamental action along with that of the quarks. The
perturbative proof of unitarity in the hadron sector rests therefore on the existence of
a consistent free action for the hadronic composites (a condition, however, which is
obviously not sufficient).

\section{A quadratic action for the pairing model}\label{sec2}

The pairing model is a simplified version of the BCS model, which describes a
system of fermions living in an energy level of given angular momentum $j$ and
interacting pairwise only when the partners are coupled to zero angular 
momentum. Such a model is a schematic representation of the physics of atomic
nuclei far from closed shells where collective excitations associated with a
spin zero composite, the Cooper pair, occur. From the point of view of field theory, 
it is a one dimensional model, the infinitely many spatial degrees freedom being
frozen to the components of the angular momentum, which can be considered as intrinsic
degrees of freedom.

The euclidean action of the pairing model defined on a lattice of spacing $a$ is
\begin{equation}
S_{\lambda}= - a\sum_{x=0}^{N-1} \sum_{m=-j}^j \left[ \overline{\lambda}_m(x) 
\nabla \lambda_m(x)\right]  - E \overline{\lambda}_m(x)\lambda_m(x-1) +g
\phi^*(x) \phi(x-1), 
\end{equation}
where 
\begin{equation}
\nabla\lambda(x)= {1 \over a }[\lambda(x)-\lambda(x-1)],
\end{equation}
$\lambda_m$ is the Grassmann variable associated to a nucleon of spin $j$
and third component of the spin equal to $m$, $E$ the nucleon excitation energy, and
g the pairing coupling. Notice that in nonrelativistic models the time
derivative is non symmetric, so that the problem of the doubling does not occur.
Finally $\phi$ and $\phi^*$ are the Cooper pair composite fields  
\begin{equation} \phi=\sum_{m={1
\over 2}}^j (-1)^{j-m}\lambda_{-m}\lambda_m,\;\;\;
\phi^*= \sum_{m={1 \over 2}}^j (-1)^{j-m} \overline{\lambda}_m 
\overline{\lambda}_{-m}, 
\end{equation}
which have index of nilpotency  $\Omega= j+1/2$. The partition function is
given by the Berezin integral
\begin{equation}
Z_{\lambda}= \int [d\overline{\lambda} d\lambda] \exp(-S_{\lambda}),
\end{equation}
where
\begin{equation}
[d\overline{\lambda} d \lambda]= \prod_{x=0}^{N-1} 
d\overline{\lambda}_{-j}(x)...d\overline{\lambda}_j(x)
           d\lambda_j(x)...d\lambda_{-j}(x)
\end{equation}
In the evaluation of the
correlation functions of the Cooper pairs we can assume $\phi$ as integration
variable replacing the Berezin integral over the nucleon field according to
\begin{equation}
\int d\lambda_j...d\lambda_{-j} \,\phi^n = \int d\phi\, \phi^n= \Omega!\,
\delta_{n,\Omega}.
\end{equation}
It can then be shown that, {\it as far as these correlation functions are
 concerned}, the partition function can be written
\begin{equation}
Z_C=\int[d\phi^*][d\phi] \exp(-S_C),\;\;\;[d\phi]=\prod_{x=0}^{N-1} d\phi(x).
\end{equation}
The Cooper pair euclidean action appearing in the above equation is
 \begin{eqnarray}
\lefteqn{  S_C= -\sum_{x=0}^{N-1} \left[g \phi^*(x)\phi(x-1) + \right.} \\
&& \left.\ln\left\{
\sum_{r=0}^{\Omega}{ 1\over {r!}}[\phi^*(x)]^r
   \sum_{s=0}^r {r\choose s}{(\Omega-s)! [\Omega -(r-s)]! \over (\Omega!)^2}
   \eta^{2(r-s)}\phi^s(x) \phi^{r-s}(x-1)\right\} \right]
\nonumber 
\end{eqnarray}
with
\begin{equation}
\eta=1-aE.
\end{equation}
Notice that the nucleon field must satisfy antiperiodic boundary conditions, so that 
the $\phi$-field satisfies the periodic ones.
We see that this action  has nothing to do with the
action of true bosons and it is static in the formal continuum limit. 

The expansion of the $\ln$ generates a polynomial of
degree $\Omega$. This polynomial has been evaluated for a few values of
$\Omega$, and it has always been found to have the form 
\begin{equation}
   S= - \sum_{n=0}^\Omega \sum_{x=0}^{N-1} \alpha_n \{[\phi^*(x)
      \phi(x)]^n+\eta^{2n}[\phi^*(x)\phi(x-1)]^n
   \}, 
\end{equation}
where the $\alpha_n $ are numerical coefficients. This form is remarkable
because it is similar to that of
the mesonic action occurring in the strong coupling limit of QCD,
characterized by the absence of terms like $(\phi^*(x))^m (\phi(x))^{m-n}
(\phi(x-1))^n,m\neq n$, due to non trivial cancellations.
 
In this paper we are only interested in the free part (obtained by putting
$g=0$), whose correlation functions in the continuum limit are  
\begin{equation}
   \lim_{a\to 0} \<[\phi^n(x_2) \phi^*(x_1)]^n\> = e^{-(t_2-t_1)n2E},
   \label{2pt} 
\end{equation}
where $t_k=ax_k$ are the physical times.
As said in the Introduction, it is natural to think that if we restrict ourselves
to the $n=1$ two-point correlation functions for the first power of the fields, we can
generate them by a much simpler action, quadratic in the fields, of the form 
\begin{equation}
   S= - \sum_{x,y=0}^{N-1} \phi^*(x) A_{x,y}\phi(y),
   \label{13}
\end{equation}
where $A$ is minus the wave operator which for the sake of convenience is parametrized
according to
\begin{equation}
A= (\alpha + \beta )  1_{} - \beta a \nabla.
\end{equation}
In the Appendix we have evaluated the  correlation functions for the above
action, with the result
\begin{equation}
    \<[ \phi^*(0) \phi(x)]^n\> = \left\{ \sum_{k=0}^\Omega f_k^N  \right\}^{-1}
    {1 \over \beta^n} \sum_{k=0}^{\Omega-n} {(k+n)!\over k!}
    (f_{k+n})^{x+1}(f_k)^{N-x-1}, \label{2pta}
\end{equation}
where
\begin{equation}
    f_k= {\Omega \choose k} \left({\beta\over 
    \alpha}\right)^k.\label{fk}
\end{equation}
To evaluate the continuum limit it is convenient to write $f_k^x$ in the form 
\begin{equation}
    f_k^x=  \left[ {1 \over {\Omega^k}} {\Omega \choose k} \right]^{ t\over a}
    e^{{kt\over a} \ln {\Omega \beta \over {\alpha}}},
\end{equation}
which immediately shows that assuming $\alpha=c\Omega, \beta=c\eta^2$ with
$c$ arbitrary, we have 
\begin{equation}
\lim_{a \to 0} f_k^x= \delta_{k,0}+\delta_{k,1} e^{-2Et}.
\end{equation}
Therefore we get the correct continuum limit for the $n=1$ two-point correlation 
functions,
while all the  correlation functions of higher powers of the composite 
vanish. 

It is also worth while noticing the form of the wave operator appearing in this
action
\begin{equation}
- A=  -[ (\Omega + \eta^2 ) 1_{} - \eta^2 a \nabla ],
\end{equation}
which is static and of sign opposite to that of a true nonrelativistic boson, 
but, as already stated, the
propagator of the Cooper pair {\em is not its inverse }.

\section{A quadratic action for the pions}\label{sec3}

The investigation of the pairing model has confirmed our conjecture about the
existence of a free action for fermionic composites, and it has also shown that 
its parameters are not related in a simple way to those of the constituents. In
particular we have seen 
the explicit appearance of the index of nilpotency 
among the parameters. In constructing an action for the pions we
will keep in mind these  results. 
The other relevant ingredient for our problem is obviously the chiral 
invariance.

The composites with the quantum numbers of the pions are
\begin{equation} 
  \vec{\pi} = i\,a^{2}\,\overline{\lambda} \gamma_5 \vec{\tau} \lambda \label{phidef}.
\end{equation}
In the above definition the $\tau_{k}$'s are the Pauli matrices and the sum over
the colour, isospin and spin  indices $a, f$ and $\beta$ of the quark field
$\lambda^{a}_{f,\beta}$  is understood. The power of the lattice 
spacing has been introduced in order to give the fields the 
canonical dimension of a boson.
To formulate a chiral invariant model we need also the field
\be
  \sigma = a^{2}\,\overline{\lambda} \lambda.
\ee 
All these composites 
have~\cite{Palu} index of nilpotency $\Omega =  24$. 
The nilpotency however is not the only feature which
distinguishes these Grassmann variables from ordinary variables. Indeed, they
are not  independent from one another, because the monomials 
\be
 \Phi_{n_{0},n_{1},n_{1},n_{3}} =  \sigma^{n_{0}} \pi_{1}^{n_{1}} \pi_{2}^{n_{2}} 
\pi_{3}^{n_{3}}
\ee  
vanish when the sum of the exponents $\sum_{k=0}^{3} n_{k}$ is larger than $\Omega$.
Moreover, if it is equal to $\Omega$, these monomials vanish if at least one of the
exponents is odd.

The chiral transformations over the quarks
\begin{eqnarray}
\delta \lambda &=& {i\over 2} \gamma_{5} \vec{\tau}\cdot \vec{\alpha} 
\lambda \\
\delta \overline{\lambda} &=& {i\over 2} \overline{\lambda} \gamma_{5} 
\vec{\tau}\cdot \vec{\alpha} 
\end{eqnarray}
induce $O(4)$-transformations over the mesons
\begin{eqnarray}
\delta \sigma &=& \vec{\alpha} \cdot \vec{\pi} \\
\delta \vec{\pi} &=& - \vec{\alpha} \sigma.
\end{eqnarray}
A quadratic action with the above symmetry must therefore
be of the form
\begin{equation}
S_{C} = -a^{-2}  \left[ {1 \over 2} (\vec{\pi},A\vec{\pi})  + {1 \over 2}
(\sigma,A\sigma)  +  (m, \sigma) \right],\label{actionC2} 
\end{equation}
where a breaking term has been included. The scalar product is defined as 
\be
(f,g) =  a^{4}\,\sum_{x} f(x) g(x).
\ee
The powers of the lattice spacing $a$ have been introduced assuming 
the wave-operator $A$ to be dimensionless.
The analogy with the model of ref.~\cite{Palu94} and the pairing model
suggest  a form of the type
\be
 A=  \rho^2 +  a^{2 }\Box ,
\ee
where $\Box$ is the Euclidean D'Alambertian on the lattice  and the parameter 
$\rho$ should be such that $A$ be positive
definite, namely $\rho^2 > 16$. We will, however, choose the more convenient 
expression
\be
 A = {\rho^4 \over - a^{2} \Box + \rho^{2}}\, 
\ee
which is equal to the previous one for small  $a^{2}$ provided that
$\rho$ is independent of $a$.
With such a definition the positivity
of $A$ requires only the positivity of $\rho^2$.

We will however consider also the case of $\rho$ and $m$ 
dependent from the index of nilpotency $\Omega$ and the lattice 
spacing $a$ according to
\begin{eqnarray}
\rho &=&  (a M_{3})^{\beta-\alpha} \label{rhop}\\
m &= & \Omega^{\gamma} (a M_{1})^{\alpha} M_{2} \label{mp}
\end{eqnarray}
where the $M_{i}$'s are independent from the lattice 
spacing and the index of nilpotency. 

The first two terms  in the 
action \reff{actionC2} have  dimension 6 and we are therefore 
free to add them to the  QCD action if $\rho$ is idependent of $a$.
The symmetry breaking  term  has instead dimension 4    (indeed it is of 
the same form of the quark mass term already
present in the  QCD action) but it formally vanishes in the continuum 
limit provided
\be
\alpha >0 \label{alphagt0}
\ee
in which case it can also be added to the QCD action.

Let us make clear that the partition function is
\begin{equation}
Z_{C} = \int [d\overline{\lambda} d\lambda]\, \exp(-S_{C}). 
\end{equation}
Here and in the following differentials appearing in square brackets are understood
as the product in a given order of the differentials over all the lattice sites and
intrinsic quantum numbers.

\section{Breaking of the chiral invariance}

In order to investigate our model we introduce the Stratonovich--Hubbard
transform~\cite{Hub} 
\begin{eqnarray}                                    
  \exp(-S_C)& = &{1\over ( \det A)^{2}} \int \left[{d(a\vec{\chi})\over\sqrt{2 \pi}}\right] 
  \left[{d(a\chi_{0})\over\sqrt{2 \pi}}\right] 
   \exp \left[ - {1\over 2 a^{2}} 
     (\vec{\chi},A^{-1}\vec{\chi}) \right.
\nonumber\\ 
  & &  \left. -{1 \over 2 a^{2}} (\chi_{0},A^{-1} \chi_{0})  +
       {1 \over  a^{2}} (\vec{\chi}, \vec{\pi}) +
       {1 \over  a^{2}} (m+\chi_{0},\sigma) \right].    
\end{eqnarray}
The partition function can then be written 
\begin{eqnarray} 
   Z_{C} &=&  {1\over ( \det A)^{2}} \int \left[{d\vec{\chi}\over\sqrt{2 \pi}}\right]  
             \left[{d\chi_{0}\over\sqrt{2 \pi}}\right] 
 \exp \left[ - {1\over 2 a^{2}} 
     (\vec{\chi},A^{-1}\vec{\chi})  -{1 \over 2 a^{2}} (\chi_{0},A^{-1} \chi_{0})  \right]   
\nonumber\\ 
& &\prod_{x}\left[D\left(\vec{\chi},\chi_{0}\right)\right]^{\Omega/2}, 
\end{eqnarray}
where
\be
   D(\vec{\chi},\chi_{0}) = a^{2} \left[ \left( m+\chi_{0} 
   \right)^{2} + \vec{\chi}^{2} \right] .
\ee
The effective action for the $\chi$ fields is therefore
\be
S_{\chi} = {1\over 2 a^{2}} \left[ (\vec{\chi},A^{-1}\vec{\chi})  
+ (\chi_{0},A^{-1} \chi_{0}) \right]  -{\Omega\over 2 a^{4}} \left(1, \ln
D\left(\vec{\chi},\chi_{0}\right) \right).
\ee
This action is $O(4)$ invariant for vanishing quark mass and reminds us of the 
Gell-Mann--L\'{e}vy model~\cite{GML}.

Since $\Omega$ is a rather large number we can  apply the saddle-point method 
and evaluate the partition function as a series in inverse powers of this
parameter. 
The saddle point equations are 
\begin{eqnarray}
 {\partial S_{\chi} \over \partial
\chi_0(x) } &=& a^{2} \left[( A^{-1}\chi_0 ) (x) - \Omega {\chi_0(x) + m \over 
D(x)} \right]=0\nonumber \\
{\partial S_{\chi} \over \partial
\chi_h(x) } &=& a^{2} \left[( A^{-1}\chi_h ) (x) - \Omega {\chi_h(x) \over 
D(x)} \right] =0. \label{saddle}
\end{eqnarray}
At constant fields the minima are characterized by the parameter 
\be
A_{0} = \sum_x A_{xy} =  \rho^{2}.
\ee 
The minimum is achieved for $\vec{\chi}=0$ and
\be
 \overline{\chi_{0}} = {-m a \pm \sqrt{m^{2}a^{2}+4 \Omega \rho^2} \over 2 a},
\ee
where the $+$ (respectively $-$) sign has to be chosen when $m>0$ 
(respectively $m<0$), which we shall assume to be the case. 

Now we assume 
 \begin{equation}
(am)^2<< 4\Omega \rho^2 \label{ine}
\end{equation}
so that 
\be
 \overline{\chi_{0}} \approx  a^{-1} [\sqrt{\Omega}  \rho - {1 \over 2} a m].  
\ee
In order to keep the two terms in the r.h.s. of the previous equation of 
the same  order in $\Omega$ we 
require for the exponent $\gamma$ that appears in \reff{mp}
\be
\gamma ={1\over 2}.
\ee
The second derivatives of $S_{\chi}$ at the minimum are
\begin{eqnarray}
\left.{\partial^2 S_\chi \over \partial \chi_0(x) \partial \chi_0(y)}
\right|_{\chi=\bar{\chi}} &=&
   a^2 \left[ A^{-1}_{xy}+ \delta_{xy}{\Omega \over { D}}  \right] 
   \nonumber \\ &=& 
  {a^4 \over {\rho^4}} \left( -\Box +  { 2 \rho^{2}\over{ a^2}} -
 m {\rho  \over a \sqrt{\Omega}}\right)  
\nonumber\\
\left.{\partial^2 S_\chi \over \partial \chi_h(x) \partial \chi_k(y)}
\right|_{\chi=\bar{\chi}} &=&
 \delta_{hk}    a^2 \left[ A^{-1}_{xy}- \delta_{xy}{\Omega \over {
D}}  \right] 
   \nonumber \\ &=& 
   \delta_{hk} {a^4 \over {\rho^4}} \left( -\Box  + m {\rho \over a \sqrt{\Omega}} 
\right).  
\end{eqnarray}
The propagator of the pion field to leading order 
\begin{eqnarray}
<\pi_h(x) \pi_k(y)> &=& {1 \over {Z_C}}  { \Omega^{2} \over {(\det 
A)^2}}\int \left[ 
{d\chi_0 \over 
\sqrt{2\pi}}\right]\left[ {d\vec{\chi} \over \sqrt{2\pi}}\right]
 {\chi_h(x)\over D(x)} 
 {\chi_k(y)\over D(y)} \exp(-S_{\chi})\nonumber \\
& =&  {1\over a^{4}} \,\left({ 1 \over { -\Box+
m_{\pi}^2}} \right)_{x,y}, \qquad \hbox{for $x\neq y$}
\end{eqnarray}
turns out to have the canonical form. Its mass is therefore
\be
m_\pi^2 = { 1 \over \sqrt{\Omega} a}   m  \rho = 
({ M_{1}\over M_{3}})^{\alpha} M_{2} M_{3} (a M_{3})^{\beta-1}. 
\label{mpi}
\ee
The mass of the $\sigma$-field is instead
\be
m_{\sigma}^2= 2 a^{-2} \rho^2 -m_{\pi}^2 = 2 a^{-2} (a 
M_{3})^{2(\beta-\alpha)} -m_{\pi}^2 . \label{ms}
\ee
As a consequence of the inequality~\reff{ine} $m_\sigma >> m_\pi$.

The $\sigma$-field  acquires the nonvanishing expectation value  
\begin{equation}
<\sigma>= {1\over V} {\partial\over \partial (a^2m)} \log Z_{C} \approx 
{a^{-2}\Omega \over \overline{\chi_{0}} +  m} \approx a^{-1} \rho^{-1}
\sqrt{\Omega}.\label{vms} 
\end{equation}
We may notice how PCAC is realized in the present model.
By using the
chiral transformations we get the relation
\be
    {\partial S_{C}\over \partial \vec{\pi}} \delta \vec{\pi} +
    {\partial S_{C}\over \partial \sigma } \delta \sigma = 0
\ee
which implies
\be
\sum_{y} \sigma(x) A_{xy} \vec{\pi}(y) -   \vec{\pi}(x) A_{xy}
\sigma(y) = m \vec{\pi}(x). \label{pcac}
\ee
In the formal continuum limit
\be
A \approx  \rho^{2} + a^{2}\Box,
\ee
 the previous identity  becomes
\be
a^{2} \left[ \sigma \Box \vec{\pi} - \vec{\pi} \Box
\sigma \right] =  m  \vec{\pi}.
\ee
The lhs of the above equation is proportional to the divergence of an
axial current, namely
\be
\vec{{\cal A}} = \sigma \nabla \vec{ \pi} - \vec{\pi}  \nabla  \sigma
\ee
so that the previous equation can  be rewritten as
\be
a^{2}  \nabla \vec {{\cal A}} =  m  \vec{\pi}.
\ee
This identity implies the relation
\be
a^2 m_\pi^2 \< \sigma \> = m
\ee

\section{The $\Omega$-expansion}

In order to formulate the $\Omega$-expansion it is convenient  to introduce the fields
which correspond to rescaled fluctuations
\begin{eqnarray}
\theta_0 & = &  \rho^{-2 }( \chi_0 - \bar{\chi}_0 ),
\nonumber\\
\theta_k & = &  \rho^{-2} \chi_k.
\end{eqnarray}
In terms of these fields the function $D$ takes the form
\be
D=  C^2 F
\ee
where
\begin{eqnarray}
C &=&\sqrt{\Omega \rho^2 + { 1\over 4} a^2m^2 } + { 1\over 2} am \approx \sqrt{\Omega}
\rho+ { 1 \over 2} am
\nonumber\\
F &=&  1+ {\rho^2 \over C} 2a \theta_0  + {\rho^4 \over C^2} a^2 ( \theta_0^2 + {\vec{\theta}}^2 ).
\end{eqnarray}
Finally the expansion of the pion propagator 
can be evaluated by the equation
\begin{eqnarray}
<\pi_h(x) \pi_k(y)> &=& { 1\over Z_C} { 1\over (\det A)^2} {\Omega^2 \rho^4 \over\
 C^4}\int
 \left[{d\theta_0 \over \sqrt{2\pi}}\right] \left[{d\vec{\theta}\over 
 \sqrt{2\pi}} \right]
\nonumber\\
  & & \cdot {\theta_h(x)\over F(x)} {\theta_k(y) \over F(y)} \exp{(-S_{\chi})}
  \quad \hbox{for $x\neq y$. }
\end{eqnarray}
By expanding the $\ln D $ we  rewrite the action $S_{\chi}$ as a series
\begin{equation}
S_{\chi}= \sum_{n=2}^{\infty}  S^{(n)}.
\end{equation}
The term $S^{(n)}$, for $n>2$, is a homogeneous polynomial of degree $n$ in the
$\theta$-fields   proportional to $ \rho^{2n} a^n / {C^n}$. The first three terms are
\begin{eqnarray}
S^{(2)} &=& a^4 \sum_x {1 \over2} \vec{\theta} \left(-\Box+m_{\pi}^2\right)
\vec{\theta}
+{1\over2} \theta_0 \left(-\Box+ m_{\sigma}^2\right) \theta_0 \nonumber\\
S^{(3)} &=&  {\rho^6 \over C^3} a^4\sum_x { 1\over a}\left(  { 2\over 3} \theta_0^3
 -2 \theta_0 \vec{\theta}^2  \right) \nonumber\\
S^{(4)} &=& { \rho^8 \over C^4} a^4\sum_x  -{1\over 2} \theta_0^4 
-{1\over 2} (\vec{\theta}^2)^2 + 3 \theta_0^2  \vec{\theta}^2 .
\end{eqnarray}
With our choices $S^{(n)}$ turns out to be of the order $\Omega^{-{n\over2}}$, for $n>2$.
We actually have then an expansion in inverse powers of $\sqrt{\Omega}$, but the first
correction is of order $\Omega^{-{3\over 2}}$. 

We can now investigate which is the dependence of $\rho$ and $m$ on the  
lattice spacing which ensures the renormalizability of the of 
$\Omega$-expansion. This requires $\rho^2 / C$ not to diverge in the
continuum limit.
Since we demand also $m_{\pi}$ to be finite we get 
from \reff{mpi}
\be
\beta =1.
\ee
As a consequence of \reff{alphagt0} the mass of the $\sigma$-field is divergent (see 
\reff{ms}). 
It follows for $C$ the following expression
\be
C = \sqrt{\Omega} (a M_{3})^{1-\alpha} 
\ee
so that for small lattice spacing
\be
{\rho^{2}\over C} \sim a^{1-\alpha}
\ee
which therefore implies $\alpha\leq 1$. 

It is remarkable that with the choice $\alpha=1/2$ we get a {\em truly 
free action} for the pions, all the interaction terms giving vanishing 
contributions to their $n$-point functions.  Such an action, however,
cannot be used to set up a new perturbative approach to QCD because it 
is an operator of dimension 6 which is not accompanied by thenecessary 
power of the lattice spacing. For this purpose one must make
a different choice of the parameters,
that is $\alpha=1$, for which
\be
 \rho\;\hbox{independent of $a$,}\;\;\;
m = { \sqrt{\Omega}\over \rho}a m_{\pi}^2,
\ee 
which provides an interacting renormalizable model. Indeed,
while for $n>4$ all the terms $S^{(n)}$ are  accompanied by a factor $a^{n-4}$ and 
are irrelevant, the terms $S^{(3)}$ and $S^{(4)}$ survive in the formal 
continuum limit. The $\theta_0$-field has a divergent mass
$ \approx 1/a$,  and it can be rescaled according to $\theta_0 \rightarrow a \theta_0$,
showing that $S^{(3)}$ does not diverge. 
The $\sigma$-field has still a divergent mass and it is unphysical.

\section{Conclusion}

Effective Grassmann actions for even composites are in general polynomials of degree
equal to the index of nilpotency of the composites. Their complexity is due to the
fact that they must generate the correlation functions of all the nonvanishing powers
of the composites. One can then think that the $n=1$ two-point functions  can be 
obtained from a much simpler action. In the case of the pairing model, where the exact 
action of the Cooper pair has been derived from that of the nucleons, we have indeed shown
that the $n=1$ correlation functions  can be generated by a quadratic action which shares
the specific features of the actions of even composites previously considered, namely
it is static in the formal continuum limit and of sign opposite to that of a true 
nonrelativistic boson. We have
then considered an action of this type for the pion. It includes the $\sigma$-field in order to be invariant 
under $[SU(2)]_L\otimes [SU(2)]_R$ transformations over the quarks. 
This model exhibits the  Goldstone phenomenon by reducing its symmetry to the 
$O(3)$ isospin invariance. In the presence of an explicit breaking linear in
$\sigma$  a squared mass for the pion is generated proportional to the 
breaking parameter.

One can regard our results as the construction of a simple model, whose main
ingredient is the compositness of the mesonic fields. This model reminds us of
that of Gell-Mann-L\'{e}vy, the
connection being very transparent after the  
Stratonovitch-Hubbard representation has been
introduced.

Our main motivation, however, is not  model building, but to use
our action to construct a new perturbative approach to QCD. Whether this
is really possible  can only be established by explicit calculations
which are now in progress.

In any case the present results complete the construction of free actions for 
composite fields, which are shown to have to large extent the properties of
elementary fields. This provides the building blocks 
for the models with quarks of lower dimensions referred to in the Introduction.

\appendix

\section{Solution of the one-dimensional model}

In this appendix we shall solve the one dimensional model defined by 
the action \reff{13} by following a renormalization-group strategy, 
that is by iterating the integration of the degrees of freedom within 
a scale length.
For our purposes the effective action at the scale $l$ is defined by
\be
e^{S_{l}} = \prod_{x}  e^{\alpha \, \phi^{*}(x) \phi(x)} I(x,l)
\ee
where the general form of the hopping part of the 
interaction between the sites $x$ and $x+l$ can be taken to be
\be
I(x,l) = \sum_{k=0}^{\Omega} C_{k}(l) \left[ \phi(x) \phi^{*}(x+l) \right]^{k}
\ee
with $x$ running on the lattice with spacing $l$.

The thinning of the degrees of freedom amounts to the evaluation of the integral
\begin{eqnarray}
I(x-l,2l) & = &  \int  d\phi^{*}(x) d\phi(x) e^{\alpha \, \phi^{*}(x) \phi(x)}
         \, I(x-l,l) I(x,l)  \nonumber \\
& = &  \int  d\phi^{*}(x) d\phi(x) \sum_{n=0}^{\Omega}{\alpha^n\over 
n!} \, \left[ \phi^{*}(x) \phi(x)\right]^{n} \cdot \nonumber \\
& & \quad \sum_{k=0}^{\Omega}
C_{k}^{2}(l) \left[ \phi^{*}(x) \phi(x) \phi(x-l) \phi^{*}(x+l) \right]^{k}
\nonumber \\
& = & \sum_{k=0}^{\Omega}{\alpha^{\Omega-k}\over 
(\Omega-k)!} \,(\Omega!)^{2}\, 
C_{k}^{2}(l) \left[ \phi(x-l) \phi^{*}(x+l) \right]^{k},  \nonumber 
\end{eqnarray}
so that
\be
C_{k}(2l) = {\alpha^{\Omega-k}\over 
(\Omega-k)!} \,(\Omega!)^{2}\, 
C_{k}^{2}(l).
\ee
By iteration we get the solution of this recursion relation 
\be
C_{k}(l) = \left[{\alpha^{\Omega-k}\over 
(\Omega-k)!} \,(\Omega!)^{2} \right]^{l-1}\ 
C_{k}^{l}(1),
\ee
which has the initial condition 
\be
C_{k}(1) = {\beta^{k}\over k!}.
\ee
By using periodic boundary conditions on a lattice of size $N$ we 
obtain for the partition function
\begin{eqnarray}
Z &=&  \int  d\phi^{*}(0) d\phi(0) e^{\alpha \, \phi^{*}(0) \phi(0)}  \, 
I(0,N)  \nonumber\\
&=& \sum_{k=0}^{\Omega} C_{k}(N) {\alpha^{\Omega-k}\over 
(\Omega-k)!} \,(\Omega!)^{2}   \nonumber\\
&=& \sum_{k=0}^{\Omega} C_{k}^{N}(1) \left[{\alpha^{\Omega-k}\over 
(\Omega-k)!} \,(\Omega!)^{2} \right]^{N}  \nonumber\\
&=& \left(\alpha^{\Omega}  \,\Omega! \right)^{N}  \sum_{k=0}^{\Omega}
 \left[\left({\beta\over \alpha}\right)^{k} {\Omega \choose k} 
 \right]^{N} \nonumber\\
&=& \left(\alpha^{\Omega}  \,\Omega! \right)^{N}  \sum_{k=0}^{\Omega} 
f_{k}^{N},
\end{eqnarray}
where the coefficients $f_{k}$ have been defined in \reff{fk}.
Similarly for the two-point functions we have
\begin{eqnarray}
\<\left[\phi^{*}(0) \phi(x)\right]^{n}\> &=& {1\over Z} 
 \int  d\phi^{*}(0) d\phi(0)  d\phi^{*}(x) d\phi(x) e^{\alpha [ 
 \phi^{*}(0) \phi(0) + \phi^{*}(x) \phi(x)]} \cdot \nonumber\\
 & & \sum_{r=0}^{\Omega} \sum_{k=0}^{\Omega-n} C_{k}(x) C_{r}(N-x) 
  \left[ \phi(0) \phi^{*}(x)\right]^{k+n}  \left[ \phi(x) 
  \phi^{*}(0)\right]^{r}\nonumber\\
&=& {1\over Z} 
 \int  d\phi^{*}(0) d\phi(0)  d\phi^{*}(x) d\phi(x) e^{\alpha [ 
 \phi^{*}(0) \phi(0) + \phi^{*}(x) \phi(x)]} \cdot \nonumber\\
 & &  \sum_{k=0}^{\Omega-n} C_{k+n}(x) C_{k}(N-x) 
  \left[ \phi^{*}(0) \phi(0) \phi^{*}(x) \phi(x)\right]^{k+n} \nonumber\\
&=& {1\over Z} \sum_{k=0}^{\Omega-n} C_{k+n}(x) 
C_{k}(N-x){\alpha^{\Omega-k-n}\over 
(\Omega-k-n)!} \,(\Omega!)^{2} \nonumber \\
&=&  \left\{ \sum_{k=0}^\Omega f_k^N  \right\}^{-1}
    {1 \over \beta^n} \sum_{k=0}^{\Omega-n} {(k+n)!\over k!}
    (f_{k+n})^{x+1}(f_k)^{N-x-1},
\end{eqnarray}
which is the result given in \reff{2pta}.

\end{document}